\documentclass[aps,pra,superscriptaddress,amsmath,amssymb,preprintnumbers,showpacs,twocolumn]{revtex4-1}

\usepackage{amssymb}
\usepackage{bm}
\usepackage{color}
\usepackage{subfigure}
\usepackage[utf8x]{inputenc}
\usepackage[english]{babel}
\usepackage[T1]{fontenc}
\usepackage{lmodern}
\usepackage{bbold}
\usepackage{amsmath}
\usepackage{amsthm}
\usepackage[pdftex]{graphicx}
\usepackage{epstopdf}
\usepackage{pgfplots}

\begin{document}

\title{Detecting initial system-environment correlations:\\ Performance of
various distance measures for quantum states}

\author{S. Wi\ss mann}
\affiliation{Physikalisches Institut, Universit\"{a}t Freiburg, Hermann-Herder-Stra\ss e 3, D-79104 Freiburg, Germany}

\author{B. Leggio}
\affiliation{Physikalisches Institut, Universit\"{a}t Freiburg, Hermann-Herder-Stra\ss e 3, D-79104 Freiburg, Germany}
\affiliation{Dipartimento di Fisica e Chimica, Universit\`{a} di Palermo, Via Archirafi 36, 90123 Palermo,
Italy}

\author{H.-P. Breuer}
\affiliation{Physikalisches Institut, Universit\"{a}t Freiburg, Hermann-Herder-Stra\ss e 3, D-79104 Freiburg, Germany}

\begin{abstract}
We study the time evolution of four distance measures in the presence of initial system-environment correlations. It is well-known that the trace distance between two quantum states of an open system may increase due to initial correlations which leads to a breakdown of the contractivity of the reduced dynamics. Here we compare and analyze, for two different models, the time evolution of the trace distance, the Bures metric, the Hellinger distance and the Jensen-Shannon divergence regarding an increase above their initial values, witnessing initial correlations. This work generalizes, deepens and corrects the study performed by Dajka \emph{et al.} [Phys. Rev. A \textbf{84} 032120 (2011)] and thereby reveals generic features of the considered distance measures with respect to the capability of detecting initial system-environment correlations.

\end{abstract}

\maketitle
\section{Introduction}\label{sec:intro}
The interaction of a physical system with an external environment is a general feature of almost any theoretical and experimental framework. In the quantum realm, the study of such an interaction and its consequences are of paramount importance in understanding the very core of quantum dynamics \cite{hpp}, its potential application to information processing \cite{niels} and its r\^{o}le in fundamental phenomena such as efficient energy transfer \cite{photo}, thermalization \cite{therm}, or quantum memory effects \cite{hpjpb}. Although the presence of an environment usually leads to decoherence of the open quantum system \cite{decoh}, in many cases the interaction with a reservoir can also enhance certain quantum features in the open system dynamics \cite{enh} and provide also some insight into environmental properties \cite{probing}. It is therefore not surprising that the analysis of open quantum system dynamics attracts a great deal of attention. In particular, many efforts have been recently devoted to the study of non-Markovian effects in the time evolution of a composite quantum system \cite{nm1,nm2}, also in connection to the presence of correlations \cite{nm3, laura1, laura2}. 

It is known that dynamical memory effects can be quantified in terms of the increases of the trace distance between two open quantum system states evolving under the same dynamical map \cite{nm1}. In addition, an increases of this distance measure above its initial value witnesses the presence of initial system-environment correlations \cite{hpbound,hpwit,prl-witness}. This property is obtained employing the subadditivity of the trace distance $D_T$ with respect to tensor products \eqref{eq:sub}, the triangle inequality \eqref{eq:triangle} and contractivity under completely positive dynamical maps \eqref{eq:contrac}, i.e. 
\begin{eqnarray}
 &D_T(\rho_1\otimes\sigma_1,\rho_2\otimes\sigma_2)\leq D_T(\rho_1,\rho_2)+D_T(\sigma_1,\sigma_2)\,,\label{eq:sub}\\
 &D_T(\rho_1,\rho_2)\leq D_T(\rho_1,\rho_3)+D_T(\rho_3,\rho_2)\,,\label{eq:triangle}\\
 &D_T(\Lambda\rho_1,\Lambda\rho_2)\leq D_T(\rho_1,\rho_2)\,.\label{eq:contrac}
\end{eqnarray}
From these relations one obtains the following upper bound for the increase of the trace distance with respect to its initial value,
\begin{align}
 D_T\bigl(&\rho_{S}^{(1)}(t),\rho_{S}^{(2)}(t)\bigr)- D_T\bigl(\rho_{S}^{(1)}(0),\rho_{S}^{(2)}(0)\bigr)\nonumber\\
 \leq&~\sum_{k=1}^2 D_T\bigl(\rho_{SE}^{(k)}(0),\rho_{S}^{(k)}(0)\otimes\rho_{E}^{(k)}(0)\bigr)\nonumber\\
&+D_T\bigl(\rho_{E}^{(1)}(0),\rho_{E}^{(2)}(0)\bigr)\,,\label{eq:bound}
\end{align}
where $\rho_{S(E)}^{(k)}(t)=\mathrm{Tr}_{E(S)}\rho_{SE}^{(k)}(t)$ denote the marginals of the total quantum states. Since $D_T\bigl(\rho_{SE},\rho_{S}\otimes\rho_{E}\bigr)$ can be interpreted as a measure for the total amount of correlations in the state $\rho_{SE}$, inequality \eqref{eq:bound} shows that an increase of the trace distance over its initial value implies that there must be either correlations in at least one initial state $\rho_{SE}^{(k)}(0)$ or the environmental states are different. The behavior of the trace distance has been extensively studied for several physical models \cite{laura1, andrea, lucka}.

Despite its operational interpretation as a measure of distinguishability, the trace distance is clearly not the only existing distance measure for two quantum states. Indeed, some comparative studies of other distance measures with respect to their sensitivity to initial system-environment correlations have been performed for some simple composite systems \cite{hanggi}. These analyses, however, represent only a first step towards a full qualitative and quantitative comparison, and the few results which are available yet do not provide a clear picture which could motivate the choice of a particular measure rather than any other. In particular, in \cite{hanggi} such a comparison has been performed for the dynamics of a two-level atom interacting with a reservoir of modes, providing for the first time an outlook of quantitative differences of four distance measures on a particular class of correlated atom-field states. In the first part of this work we correct, deepen and generalize the analysis carried out in Ref.~\cite{hanggi}. To this end we perform, for the special class of quantum dynamics for a two-level system known as dephasing process \cite{hpp,dephasing}, an extended numerical comparison of the sensitivity to initial correlations of the four distance measures treated in Ref.~\cite{hanggi}, which gives strong numerical evidences of two well-separated classes of behaviors. We stress here that these two classes have not been detected in \cite{hanggi} where, on the contrary, the authors claim that the four considered measures show qualitatively the same sensitivity.

The paper is structured as follows: In Sec.~\ref{sec:system} we introduce the physical system and define the four measures of interest. The numerical comparison of these measures is presented in Sec.~\ref{sec:comparison}, where the main results of the paper are shown. These results are discussed in Sec.~\ref{sec:ent}, where we provide a physical picture, studying the time evolution of the atom-field entanglement. In Sec.~\ref{sec:spinstar} we perform a similar numerical analysis on a two-level system with a spin bath which confirms and strengthens our previous results. Finally, some remarks and conclusions are drawn in Sec.~\ref{sec:conc}.

\section{Physical model and distance measures}\label{sec:system}
In this section we analyze an open quantum two-level system which interacts with a finite bosonic environment. The spin system, which can be given for example by two internal degrees of freedom of an atom, is coupled to a single mode electromagnetic field yielding a pure dephasing process of the open system. Related models have been extensively studied in the last years \cite{lucka,deph1} thanks to their relative simplicity, the strong experimental connection and their rich variety of behaviors. The complete Hamiltonian of these kind of systems is given by
\begin{equation}\label{eq:hamiltonian}
 H=H_S\otimes\mathbb{1}_E+\mathbb{1}_S\otimes H_E+\sigma_z\otimes H_I\,,
\end{equation}
where $H_S=\epsilon\sigma_z$ is the free Hamiltonian of the two-level system and the environmental Hamiltonian satisfies $H_E=\omega a^{\dag}a$. The interaction term yielding the dephasing is given by $H_I=g_0(a+a^{\dag})$, where $g_0$ determines the coupling strength between system and environment.\\
In Ref.~\cite{hanggi}, this model has been used to perform a comparative study of different distance measures in terms of their qualitative behavior in the presence of initial correlations. Starting from a total initial state (atom+field) of the form
\begin{equation}\label{eq:initstatehan}
 |\Psi^{\lambda}(0)\rangle=b_1|e\rangle\otimes|0\rangle+b_2|g\rangle\otimes|\Omega_{\lambda}\rangle\,,
\end{equation}
where $|e\rangle$ ($|g\rangle$) refers to the excited (ground) state of the two-level system, $|\Omega_{\lambda}\rangle=C_{\lambda}^{-1}\left\{(1-\lambda)|0\rangle+\lambda|z\rangle\right\}$ is a field state given as a coherent superposition of the vacuum state $|0\rangle$ and a generic field state $|z\rangle$, $\lambda\in[0,1]$ and $C_{\lambda}=\sqrt{(1-\lambda)^2+\lambda^2+2\lambda(1-\lambda)\mathrm{Re}\langle0|z\rangle}$, the authors of Ref.~\cite{hanggi} compared and analyzed the capability of an increase above the initial value of different distance measures for the corresponding open system states. Such an increase is a witness for initial system-environment correlations. Due to the very definition of $|\Psi^{\lambda}(0)\rangle$, the constant $\lambda$ plays the rôle of a correlation parameter as any nonzero value of $\lambda$ results in an entangled and, therefore, correlated total initial state. Moreover, it can be shown that the initial correlations are a monotonically increasing function of $\lambda$ with respect to some quantifier of bipartite entanglement (see Sec.~\ref{sec:ent}). Our aim is to extend and generalize the study presented in Ref.~\cite{hanggi} in order to provide more physical insight into the various behaviors of the different distance measures.

To this end, we consider the following generalized class of initial states for the total system defined by
\begin{equation}\label{eq:initstate}\begin{split}
 |\Psi^{\lambda}_{U}(0)\rangle&=b_1\big(u_{11}|e\rangle+u_{21}|g\rangle\big)\otimes|0\rangle\\
&\,+b_2\big(u_{12}|e\rangle+u_{22}|g\rangle\big)\otimes|\Omega_{\lambda}\rangle\,.
\end{split}\end{equation}
States of this class are obtained applying a generic local unitary transformation $U=\mathcal{U}\otimes\mathbb{1}_E$ with $\mathcal{U}=\bigl(\begin{smallmatrix}u_{11}&u_{12}\\u_{21}&u_{22}\end{smallmatrix}\bigr)$ to the state $|\Psi^{\lambda}(0)\rangle$ in Eq.~\eqref{eq:initstatehan}. We assume that the unitary time evolution of these states is still characterized by the Hamiltonian \eqref{eq:hamiltonian}, so that the reduced state of the open system at time $t$, which are obtained by tracing out the environmental degrees of freedom of the states $|\Psi^{\lambda}_{U}(0)\rangle$, obeys
\begin{equation}\label{eq:systemstate}
 \rho_S^{\lambda}(t)=		    \begin{pmatrix}
				      p^{e}	&		B_U^{\lambda}(t)	\\
				      B_U^{\lambda}(t)^*	&		1-p^{e}
		   \end{pmatrix}\,,
\end{equation}
when the eigenstates of $\sigma_z$ are identified with the standard basis of $\mathbb{R}^2$. The coherence factor $B_U^{\lambda}(t)$ is specified by the correlation parameter $\lambda$, the local unitary $\mathcal{U}$, the weights of the coherent superposition $b_1$ and $b_2$ and the generic field state $|z\rangle$. More precisely, for a coherent state $|z\rangle$ of the bosonic field, one obtains
\begin{align}\label{eq:coherence factor}
 B_U^{\lambda}(t)=&\langle -\alpha(t)|\alpha(t)\rangle\cdot \Bigl\{|b_1|^2 u_{11}u^*_{21}+\bigl(|b_2|C_\lambda^{-1}(1-\lambda)\bigr)^2\nonumber\\
&\cdot u_{12}u^*_{22}+C_\lambda^{-1}(1-\lambda)\bigl[u_{11}u^*_{22}b_1 b^*_2+u_{12}u^*_{21}b^*_1 b_2\bigr]\Bigr\}\nonumber\\
&+\langle -\alpha(t)|z+\alpha(t)\rangle\cdot\lambda C_\lambda^{-1} A(t) u_{12}\nonumber\\
&\cdot\Bigl\{|b_2|^2 C_\lambda^{-1}(1-\lambda)u^*_{22}+b^*_1 b_2 u^*_{21}\Bigr\}\nonumber\\
&+\langle z-\alpha(t)|z+\alpha(t)\rangle\cdot\bigl(\lambda C_\lambda^{-1} A(t)|b_2|\bigr)^2 u_{12}u^*_{22}\nonumber\\
&+\langle z-\alpha(t)|\alpha(t)\rangle\cdot\lambda C_\lambda^{-1} A(t)u^*_{22}\nonumber\\
&\cdot\Bigl\{|b_2|^2 C_\lambda^{-1}(1-\lambda)u_{12}+b_1 b^*_2 u_{11} \Bigr\}\,,
\end{align}
where $\alpha(t)=g_0\omega^{-1}(1-e^{i\omega t})$, $A(t)=e^{\tfrac{1}{2}(\alpha(t)z^*-\alpha^*(t)z)}$ and $\langle x|y\rangle=\mathrm{Exp}[\tfrac{1}{2}(|x|^2 + |y|^2 - 2 x^*y)]$ denotes the overlap of two coherent states of the bosonic field. The populations $p^{e}$ are time-independent since the system undergoes a pure dephasing process and are given by
\begin{align}\label{eq:pop}
 p^e=&|b_1|^2|u_{11}|^2 +|b_1|^2|u_{12}|^2 C_\lambda^{-2}\nonumber\\
&\cdot\Bigl\{(1 -\lambda)^2  + \lambda^2+2\lambda(1-\lambda)\mathrm{Re}\langle z|0\rangle\Bigr\} \nonumber\\
&+2C_\lambda^{-1} \mathrm{Re}\big[b_1b_2^*u_{11} u_{12}^*(1-\lambda+\lambda\langle z|0\rangle)\bigr]\,.
\end{align}

As mentioned previously, a nonzero value of $\lambda$ in \eqref{eq:initstate} results in an entangled total state so that the state of the open system is mixed. Our further analysis focuses on distance measures applied to the reduced states \eqref{eq:systemstate}. The reduced open system states are clearly functions of both time and $\lambda$ so that the considered measures will also have this dependence. This is particularly convenient since we can thus control the strength of the initial system-environment correlations by tuning $\lambda$.

The goal of our analysis is to study the sensitivity of different distance measures regarding initial system-environment correlations in the reduced system dynamics. These correlations may be witnessed by an increase of the distance between two open system states above their initial value \cite{hpbound}. Following Ref.~\cite{hanggi}, we quantify the distance between two generic states $\rho_1$ and $\rho_2$ by four different measures \cite{niels, bengst, hayashi}:\\ \\
The \emph{trace distance}
\begin{equation}\label{trace}
 D_T(\rho_1,\rho_2)=\frac{1}{2}\mathrm{Tr}\sqrt{(\rho_1-\rho_2)^2}\,,
\end{equation}\\
the \emph{Bures metric}
\begin{equation}\label{bures}
 D_B(\rho_1,\rho_2)=\sqrt{1-\sqrt{F(\rho_1,\rho_2)}}\,,
\end{equation}\\
the \emph{Hellinger distance}
\begin{equation}\label{hellinger}
 D_H(\rho_1,\rho_2)=\sqrt{1-\mathrm{Tr}\big(\sqrt{\rho_2}\sqrt{\rho_1}\big)}\,,
\end{equation}\\
and the \emph{Jensen-Shannon divergence}
\begin{equation}\label{jensen}
 D_J(\rho_1,\rho_2)=\sqrt{S\Big(\frac{\rho_1+\rho_2}{2}\Big)-\frac{1}{2}S(\rho_1)-\frac{1}{2}S(\rho_1)}\,,
\end{equation}
where $F(\rho_1,\rho_2)=\big(\mathrm{Tr}\sqrt{\sqrt{\rho_2}\rho_1\sqrt{\rho_2}}\big)^2$ denotes the \emph{fidelity} of two quantum states and $S(\rho)=-\mathrm{Tr}\rho\ln\rho$ is the well-known \emph{von Neumann entropy}. We normalize all distance measures so that they obey $0\leq D_k(\rho_1,\rho_2)\leq1$ ($k=T,B,H,J$) for all quantum states $\rho_1$ and $\rho_2$.

In what follows, we fix the energy splitting of the system and the frequency of the bosonic bath as well as the coupling strength of the Hamiltonian \eqref{eq:hamiltonian} to $\epsilon=1$, $\omega=1$ and $g_0=0.1$\,. Moreover, we choose a coherent state $|z\rangle$ with $z=1$ for the field state present in $|\Omega_\lambda\rangle$.

\section{Comparison of different measures}\label{sec:comparison}

\begin{figure}[pt!]
   \centering
		\hspace{-6cm}a)\\\includegraphics[width=0.33\textwidth]{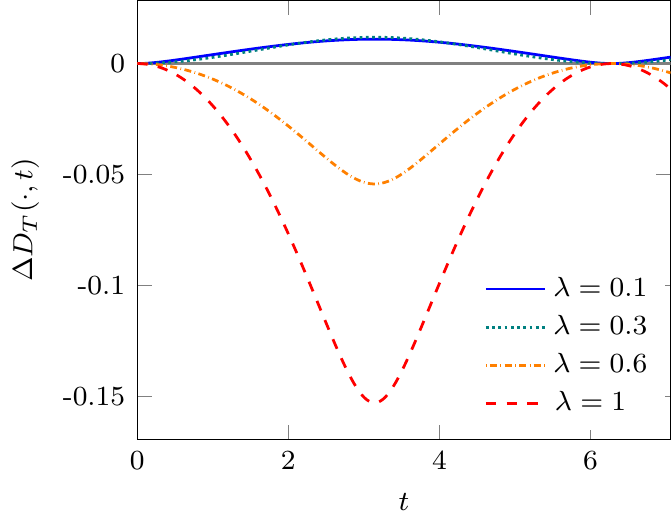}\\
    		\hspace{-6cm}b)\\\includegraphics[width=0.33\textwidth]{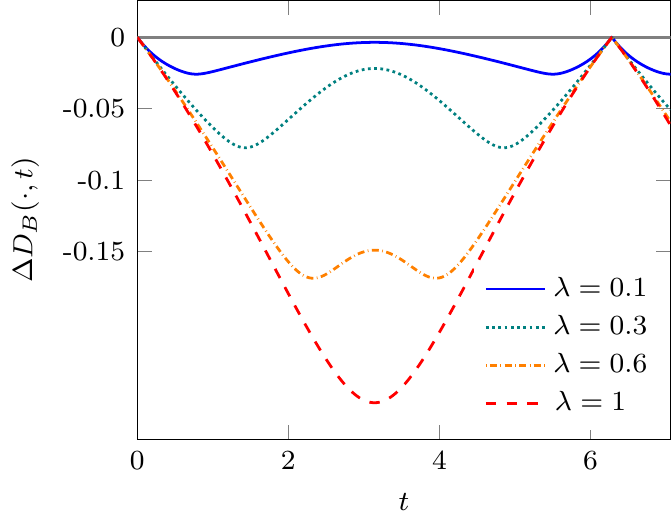}\\
  		\hspace{-6cm}c)\\\includegraphics[width=0.33\textwidth]{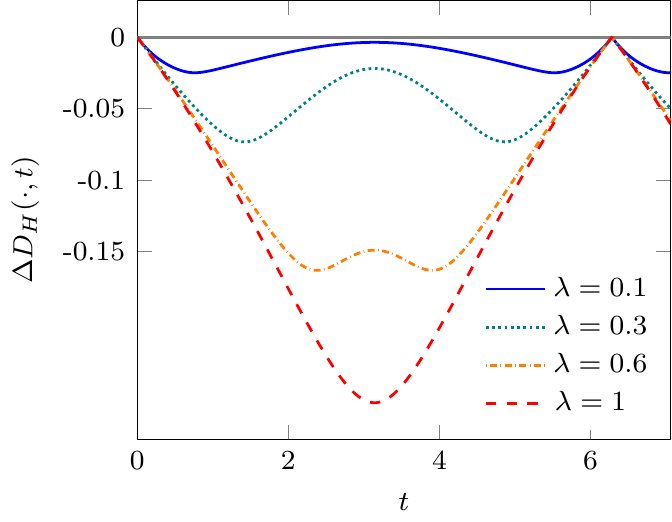}\\
    		\hspace{-6cm}d)\\\includegraphics[width=0.33\textwidth]{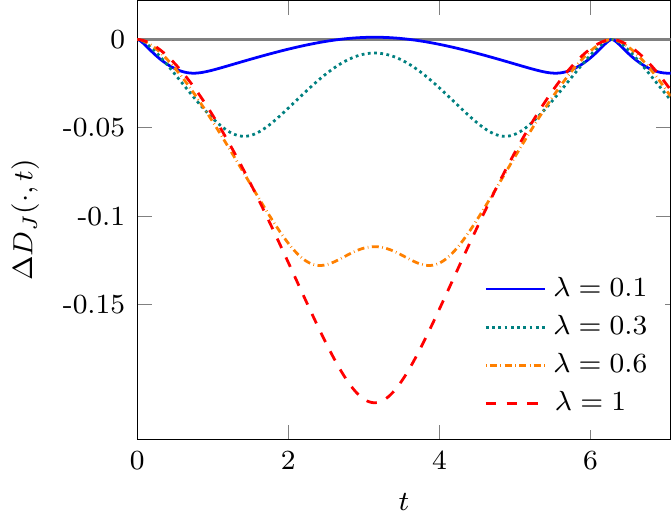}	    
    \caption{(Color online) Plot of $\Delta D_k(\lambda,t)$ for $k=T,B,H,J$ (from top to bottom; a)-d)) as a function of time $t$ for different values of the correlation parameter $\lambda$. The state of the open system is given by the marginal of the time evolved pure state \eqref{eq:initstatehan} with $b_1=b_2=1/\sqrt{2}$.}\label{fig:trace-jensen}
 \end{figure}

In order to compare the different sensitivities regarding initial system-environment correlations, we compute for each of the measures \eqref{trace}-\eqref{jensen} the change over time of the distance between an uncorrelated ($\lambda=0$) and a $\lambda$-correlated reduced state. That is, we consider
\begin{equation}\label{var}
 \Delta D_k(\lambda,t)=D_k\big(\rho_S^{\lambda}(t),\rho_S^{0}(t)\big)-D_k\big(\rho_S^{\lambda}(0),\rho_S^{0}(0)\big)\,,
\end{equation}
for $k=T,B,H,J$\,. If this quantity is positive for some time $t$ and (nonzero) value of $\lambda$, the corresponding measure is able to reveal initial correlations in the total state $|\Psi^{\lambda}_{U}(0)\rangle$ in the relative evolution of $\rho_S^{\lambda}$ and $\rho_S^{0}$. 

As a first subclass of initial states we consider those analyzed in Ref.~\cite{hanggi}. The open system state is thus given by the marginal of the time evolved pure state \eqref{eq:initstatehan}, $\rho_S^\lambda(t)=\mathrm{Tr}_E |\Psi^{\lambda}(t)\rangle\langle\Psi^{\lambda}(t)|$($=\mathrm{Tr}_E |\Psi_\mathbb{1}^{\lambda}(t)\rangle\langle\Psi_\mathbb{1}^{\lambda}(t)|$), with $b_1=b_2=1/\sqrt{2}$. Fig.~\ref{fig:trace-jensen} shows the time evolution of the four quantities $\Delta D_k(\lambda,t)$ for some special values of $\lambda$. Note that, due to the finiteness of the environment, the implemented atomic dynamics is periodic. Thanks to our particular choice of the Hamiltonian parameters the period of the dynamics is $2\pi$. The time evolutions of the distance measures presented in Fig.~\ref{fig:trace-jensen} are different from the plots given in Ref.~\cite{hanggi}, where the authors seem to have plotted the absolute value of $\Delta D_k(\lambda,t)$ which does not provide any detailed information on the capability of witnessing initial correlations.

Some important features of the plots in Fig.~\ref{fig:trace-jensen} are worth noticing: the Bures metric and the Hellinger distance never increase above their initial values, i.e. $\Delta D_{B,H}(\lambda,t)\leq0$ $\forall\,\lambda,t$ and are thus not capable to detect the initial system-environment correlations present in one of the considered states. The same holds true for large values of $\lambda$ for the two other distance measures but, for sufficiently small values of $\lambda$, the trace distance and the Jensen-Shannon divergence show an increase above their initial values. The transition regarding the capability of detecting the initial correlations takes place at $\lambda_{\text{crit.}}^T\approx 0.4$ and $\lambda_{\text{crit.}}^J\approx 0.2$ for the trace distance and the Jensen-Shannon divergence, respectively.

\begin{figure}[!t]
   \centering
	    	\includegraphics[width=0.475\textwidth]{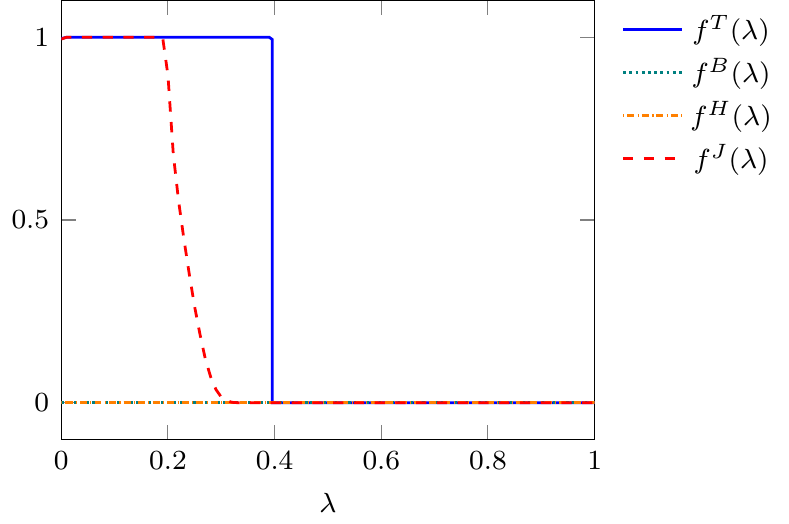}\\
   \caption{(Color online) Plot of the probability of an increase $f^k(\lambda)$ ($k=T,B,H,J$) above the initial value of the considered distance measures for reduced states given by the marginals of the states $|\Psi^{\lambda}(t)\rangle$ (cf. Eq.~\eqref{eq:initstatehan}) and randomly sampled superposition amplitudes $b_1$ and $b_2$. The lines show the fraction of states in the simulations, having a positive $\Delta D_k(\lambda, t)$ for each fixed $\lambda$ and for at least one $t \in [0,2\pi)$.}\label{fig:allsampling}
 \end{figure}

In order to extend the study of this special behavior of the different distance measures and to reveal some generic features, we performed a numerical analysis on the same class of states (cf. Eq.~\eqref{eq:initstatehan}). Sampling randomly over many different pairs of amplitudes $\{b_1,b_2\}\in\mathbb{C}^2$ such that $|b_1|^2+|b_2|^2=1$, we checked for universality in the time evolutions related solely to the general structure of the initial states rather than to the particular values of the superposition amplitudes. We generated $5\cdot10^4$ pairs of random values $\{b_1,b_2\}$ and investigated the occurrence of a positive $\Delta D_k(\lambda,t)$ for each of the corresponding states. This means that the specific distance measure increases at least for a single time interval $[t_i^k,t_f^k]$ above its initial value for a given value of $\lambda$. Fig.~\ref{fig:allsampling} shows the relative frequency $f^k$ of the occurrence of an increase as a function of the correlation parameter $\lambda$. The Bures metric and the Hellinger distance have a zero frequency of increase for all $\lambda$. Thus, there is strong numerical evidence that these two distance measures generally cannot detect the initial system-environment correlations present in any state of the form of equation \eqref{eq:initstatehan} and for any value of $\lambda$. On the other hand, the trace distance and the Jensen-Shannon divergence show a transition in the capability of witnessing initial correlations. While an increases of the trace distance and the Jensen-Shannon divergence above their respective initial value is very likely for $\lambda\ll1$, these two measures cannot detect the initial correlations for large values of $\lambda$ although the strength of the initial correlations is a monotonically increasing function of $\lambda$, as mentioned previously. Hence, only sufficiently \emph{weak} initial system-environment correlations in states of type \eqref{eq:initstatehan} can be witnessed by these two distance measures. The transition from $f^{T,J}\approx1$ to $f^{T,J}=0$ is very smooth for the Jensen-Shannon distance while $f^T$ shows a relatively sharp transition at $\lambda_{\text{trans}}^T\approx 0.4$ which is very close to the value found previously for the choices of superposition amplitudes $b_1=b_2=1/\sqrt{2}$\,. Hence, there is numerical evidence that the capability of the trace distance to detect initial correlations is almost unaffected by the choice of the superposition amplitudes and, therefore, the threshold in $\lambda$ resembles a generic feature solely related to the general structure of the initial state. A physical interpretation of this threshold is discussed later on in Sec.~\ref{sec:ent}. 

We want to point out that our findings contradict the claim stated in Ref.~\cite{hanggi} that all distance measures show an increase above their initial values for all $\lambda$ in the case of a finite-dimensional environment. The respective time evolution of the different distance measures displayed in Fig.~\ref{fig:trace-jensen} show the correct behavior and conflict those presented in Ref.~\cite{hanggi}. Moreover, we stress that neither the lower nor the upper bound for the change $\Delta D_T(t^{\prime},t,\rho_{1,2})$ of the trace distance of two states $\rho_1$ and $\rho_2$ at times $t$ and $t^{\prime}$ given in \cite{laura1,laura2,andrea} can fully explain the observed behavior of $\Delta D_T(\lambda,t)$. Only for small values of $\lambda$ these criteria provide some information about the possible increase of the trace distance above the initial value.

\begin{figure}[!t]
   \centering
		\hspace{-8.25cm}a)\\\includegraphics[width=0.475\textwidth]{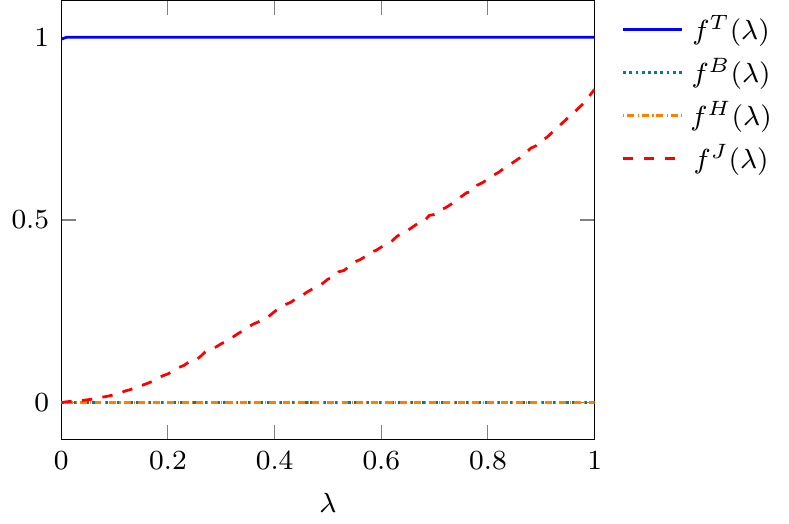}\\
  		\hspace{-8.25cm}b)\\\includegraphics[width=0.475\textwidth]{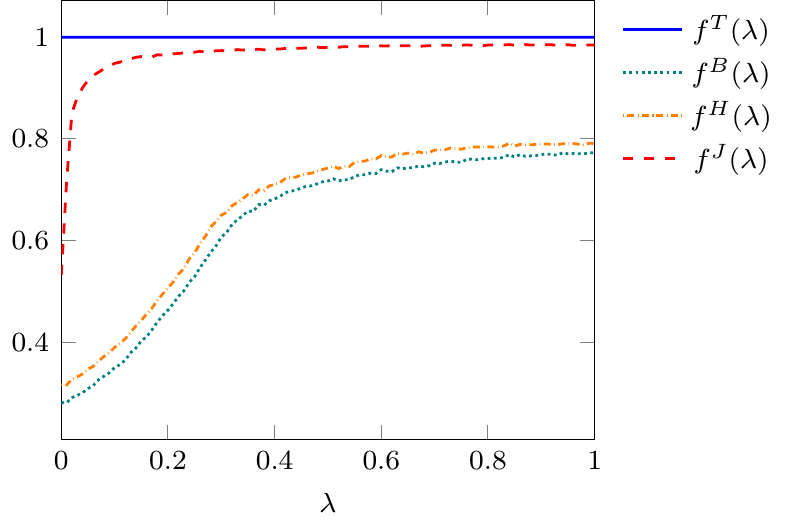}\\    
	    	\hspace{-8.25cm}c)\\\includegraphics[width=0.475\textwidth]{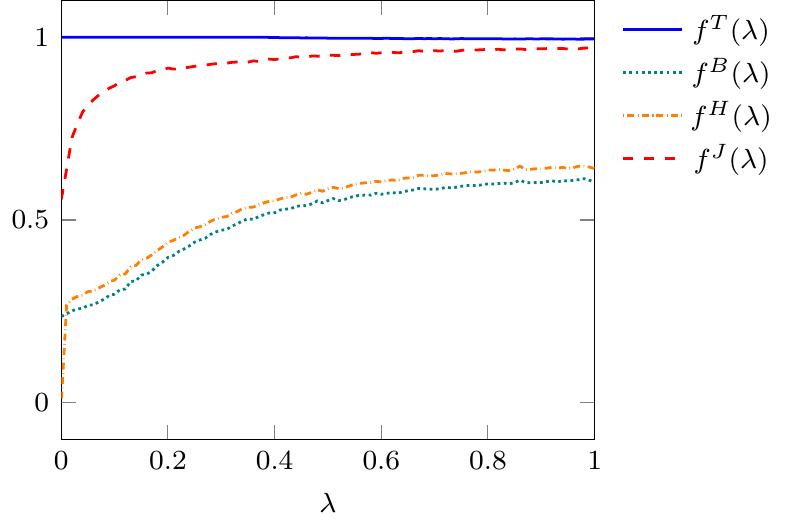}
    \caption{(Color online) Plot of the frequency of increase $f^k(\lambda)$ ($k=T,B,H,J$) of the considered distance measures for the marginals of the states $|\Psi^{\lambda}_{s}(t)\rangle$ (a)), $|\Psi^{\lambda}_{x}(t)\rangle$ (b))and $|\Psi^{\lambda}_{U}(t)\rangle$ (c)) (cf. Eq.~\eqref{eq:initstateswap}, \eqref{eq:initstatesigmax} and \eqref{eq:initstate}, respectively) with randomly sampled superposition amplitudes $b_{1,2}$. For $|\Psi^{\lambda}_{U}(t)\rangle$ the sampling is extended to local unitaries $U$ distributed with respect to the Haar measure.}\label{fig:unisampling}
 \end{figure}

In the rest of this section we generalize our studies and investigate whether the particular behavior found for states given by \eqref{eq:initstatehan} (cf. Fig.~\ref{fig:allsampling}) are maintained also for different types of initial states. We thus extend our simulations to the following two classes of total initial states,
\begin{equation}\label{eq:initstateswap}
 |\Psi^{\lambda}_s(0)\rangle=b_1|e\rangle\otimes|\Omega_{\lambda}\rangle+b_2|g\rangle\otimes|0\rangle\,,
\end{equation}
and
\begin{equation}\label{eq:initstatesigmax}
 |\Psi^{\lambda}_x(0)\rangle=b_1|1_x\rangle\otimes|\Omega_{\lambda}\rangle+b_2|-1_x\rangle\otimes|0\rangle\,,
\end{equation}
which are obtained from Eq.~\eqref{eq:initstate} for special choices of local unitaries $U$. Here, $|\pm1_x\rangle$ refers to the eigenstate of $\sigma_x$ corresponding to the eigenvalue $\pm1$. States given by Eq.~\eqref{eq:initstateswap} are obtained from Eq.~\eqref{eq:initstatehan} by a swap of the field states, so that the two vectors ($|e\rangle\otimes|\Omega_{\lambda}\rangle$ and $|g\rangle\otimes|0\rangle$), which are involved in the superposition, are characterized by very different mean values of the total energy. The second class of states (cf. Eq.~\eqref{eq:initstatesigmax}) describes, on the other hand, a different initial preparation of the total state in which entanglement is created between the system and the environment by energy exchange with the $\sigma_x$ degrees of freedom of the spin system. For both classes we performed a simulation analogous to the one reported for the states \eqref{eq:initstatehan}: we sampled $5\cdot10^4$ random values for $b_1$ and $b_2$, satisfying the normalization $|b_1|^2+|b_2|^2=1$, and checked for the occurrence of a positive value of $\Delta D_k(\lambda,t)$ for some instants of time for any given value of $\lambda$. The results of these simulations are shown in the two upper plots, a) and b), of Fig.~\ref{fig:unisampling}\,.

Some interesting features appear for the considered types of total initial states. For the \emph{swapped} states $|\Psi^{\lambda}_s\rangle$ (cf. Fig.~\ref{fig:unisampling} a)), there is no $\lambda$-threshold for the trace distance and, in addition, the probability of increase $f^T$ is always approximately unity whereas the Bures metric and the Hellinger distance show again a constant, vanishing frequency of increase and are thus also insensitive to initial correlations in this setup. The Jensen-Shannon divergence, on the other hand, exhibits a monotonic increase of its capability of witnessing the initial correlations, starting from almost vanishing frequency of increase, which grows almost linearly with $\lambda$.  

On the other hand, for the states in the $\sigma_x$-eigenbasis all distance measures are able to reveal the initial correlations in the total state $|\Psi^{\lambda}_x(0)\rangle$ (cf. Fig.~\ref{fig:unisampling} b)). However, the relative frequency of increase of the trace distance and the Jensen-Shannon divergence is significantly higher for all values of $\lambda$ than the one corresponding to the Bures metric and the Hellinger distance. The Bures metric has the smallest probability to reveal the system-environment correlations in the total initial state but its frequency of increase grows monotonically with $\lambda$. The Hellinger distance shows the same functional dependence with respect to $\lambda$, emphasizing that the four distance measure group into pairs: our results suggest that the trace distance and the Jensen-Shannon divergence are strongly connected regarding the capability of witnessing initial correlations and the same holds for the two other measures. The similarity between the Bures metric and the Hellinger distance is actually not so surprising as both measures depend essentially on the product of the square roots of the two input states (cf. \eqref{bures} and \eqref{hellinger}).

Finally, in the lowermost plot, Fig.~\ref{fig:unisampling} c), we present the results of the numerical simulation for the most general total initial states given by Eq.~\eqref{eq:initstate} with arbitrary weights $\{b_1,b_2\}$ and local unitaries $U$. In this case, we sampled not only random values for the superposition amplitudes but also random local unitaries $U$, distributed with respect to the Haar measure, and determined a potential increase of these random states. The sample size for this simulation is again $5\cdot10^4$. The considered distance measures again group into pairs and the probability of witnessing the correlations present in the total initial state is significantly higher for the trace distance and the Jensen-Shannon entropy than for the Bures metric and the Hellinger distance. While the probability of an increase above the initial value as a function of $\lambda$ is again almost unity for the trace distance, the frequency of increase for the Jensen-Shannon divergence increases again monotonically with $\lambda$ approaching unity. Moreover, one clearly observes the similarity between Fig.~\ref{fig:unisampling} b) and c). We conclude from this that the states in the $\sigma_x$-eigenbasis given by \eqref{eq:initstatesigmax} describe very well the qualitative behavior of the general class of states $|\Psi^{\lambda}_{U}(0)\rangle$ on average, and the firstly considered states \eqref{eq:initstatehan} and \eqref{eq:initstateswap} are exceptional. Moreover, we infer that the particular features of the different distance measures are not related to a special choice of $U$ rather than to the implementation of the correlations, that is, the structure of the environmental states in the total initial state.

The fact that the frequency of increase of the trace distance $f^T(\lambda)$ is mostly considerably larger than that of the Bures metric $f^B(\lambda)$ is somehow surprising. As shown previously, the trace distance is a witness for initial system-environment correlations \cite{hpbound, hpwit} due to inequality \eqref{eq:bound}. We show in the following that the Bures metric \eqref{bures} obeys the same inequality, so that these two distance measures share indeed the same features. 
The first two properties needed to derive inequality \eqref{eq:bound}, the triangle inequality \eqref{eq:triangle} and the contractivity under CPT-maps \eqref{eq:contrac}, are clearly satisfied by the Bures metric. The last property, which is required in order to show that a distance measure is a witness for initial system-environment correlations, is the subadditivity, i.e.
\begin{equation}\label{subad}
D_B(\rho_1\otimes\sigma_1,\rho_2\otimes\sigma_2)\leq D_B(\rho_1,\rho_2)+D_B(\sigma_1,\sigma_2)\,.
\end{equation}
In order to show this inequality for the Bures metric one considers the real function 
\begin{equation}\label{subrs}
Q(R,S)=\sqrt{1-\sqrt{R}}+\sqrt{1-\sqrt{S}}-\sqrt{1-\sqrt{RS}}\,,
\end{equation}
where $R, S$ are real-valued and satisfy $0\leq R,S\leq1$. It is straightforward to prove that $Q(R,S)\geq 0$ for all allowed choices of $R$ and $S$. Since the fidelity $F$ satisfies $0\leq F(\rho_1,\rho_2)\leq 1$ for any two states $\rho_1$, $\rho_2$  by its very definition, one can identify the real numbers $R$ and $S$ with the fidelity of some specific pairs of states, i.e. $R=F(\rho_1,\rho_2)$ and $S=F(\sigma_1,\sigma_2)$. In this way, one obtains a non-negative function $\mathcal{Q}(\rho_{1,2},\sigma_{1,2})\geq0$ of any four quantum states. This condition, however, can be rewritten as
\begin{equation}\begin{split}\label{subadfid}
 &\sqrt{1-\sqrt{F(\rho_1,\rho_2)F(\sigma_1,\sigma_2)}}\\
&\leq\sqrt{1-\sqrt{F(\rho_1,\rho_2)}}+\sqrt{1-\sqrt{F(\sigma_1,\sigma_2)}}\,.
\end{split}\end{equation}
Exploiting the well-known property
\begin{equation}
 F(\rho_1\otimes\sigma_1, \rho_2\otimes\sigma_2)=F(\rho_1,\rho_2)F(\sigma_1,\sigma_2)\,,
\end{equation}
of the fidelity and the definition of $D_B$ \eqref{bures}, one easily sees that Eq.~\eqref{subadfid} is equivalent to \eqref{subad} which thus holds for any choice of quantum states. Hence, the Bures metric is subadditive and, therefore, also a proper witness for initial correlations in the sense of Refs.~\cite{hpbound,hpwit}, like the trace distance. Nevertheless, our previously presented results show that (at least for the particular model considered in this work) the sensitivity of the Bures metric regarding initial system-environment correlations is much weaker than the one of the trace distance.

\section{Correlations and trace distance: a physical picture}\label{sec:ent}
Out of the two distance measures which are proven to be witnesses for initial system-environment correlations, the trace distance is, as shown in Sec.~\ref{sec:comparison}, more sensitive for the considered model. In addition, it also shows a much richer structure as a function of $\lambda$ and time. The most prominent feature of the trace distance is the threshold in $\lambda$ (cf. Fig.~\ref{fig:allsampling}), sharply separating two phases in the time evolution. Although this behavior is generic for the entire class of initial states of type \eqref{eq:initstatehan} we focus now on the firstly considered class of states with equal weights $b_1=b_2=1/\sqrt{2}$ and turn to a physical analysis of this features. 

\begin{figure}[!t]
   \centering
	    	\includegraphics[width=0.4\textwidth]{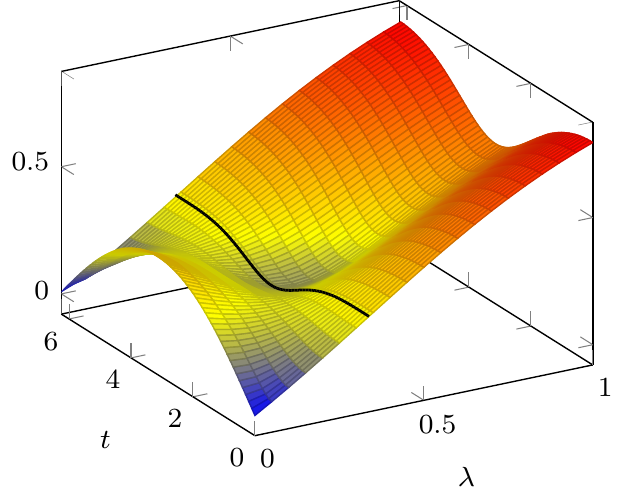}
    \caption{(Color online) Plot of the concurrence $C(|\Psi^{\lambda}(t)\rangle)$ as a function of the correlation parameter $\lambda$ and time $t$. The black line indicates the transition where the concurrence does not increase above its initial value anymore.}\label{fig:concurrence}
 \end{figure}

As mentioned before, initial correlations in the states \eqref{eq:initstatehan} are a monotonically increasing function of $\lambda$. This can be shown by evaluating the concurrence $C(\rho_{SE})$ \cite{conc,conc2} of the total system state defined by
\begin{equation}
 C(\rho_{SE})=\sqrt{2\big(1-P(\rho_S)\big)}\,,
\end{equation}
where $P(\rho_S)=\mathrm{Tr}\rho^2_S$ is the purity of the reduced system
state $\rho_S=\mathrm{Tr}_E\rho_{SE}$.
Hence, the concurrence vanishes for any pure reduced states and is equal to unity for any maximally mixed reduced states which correspond to factorizing and maximally entangled total states, respectively. Fig.~\ref{fig:concurrence} shows the quantity $C(\lambda,t)\equiv C(|\Psi^{\lambda}(t)\rangle)$ as a function of $\lambda$ and $t$, where $|\Psi^{\lambda}(0)\rangle$ is given by \eqref{eq:initstatehan} with $b_1=b_2=1/\sqrt{2}$. It can be readily seen that the initial value of the concurrence ($t=0$) increases monotonically with $\lambda$. Moreover, the concurrence has a threshold in $\lambda$ similar and, in addition, close to the one found for the trace distance $D_T$ for this class of states. The change of behavior visualized by the black line in Fig.~\ref{fig:concurrence} is characterized by the transition from a monotonically increasing concurrence from time zero to $t=\pi$ to a monotonically decreasing evolution of the concurrence. That is, entanglement quantified by $C$ increases in time up to $t=\pi$ only for small values of $\lambda$ ($\lambda_{\text{crit.}}\approx 0.34$), while for higher values of $\lambda$ the entanglement between system and environment cannot be enhanced during the evolution of time, i.e. the concurrence decreases. 

This transition from an increasing to a decreasing time evolution of the concurrence may be explained in terms of the entangling power of the unitary dynamics: roughly speaking, the higher the initial correlations in the total state are, the less probable it is for the unitary total dynamics to create even more entanglement. This is due to the fact that it is less probable for the target state of the evolution to be more entangled than the initial one. From Fig.~\ref{fig:concurrence} one can see that the highest amount of correlations created by the dynamics, i.e. the maximal value of the quantity $C(\lambda,t)-C(\lambda,0))$ for $t\in[0,\pi]$, is achieved when $\lambda$ is zero which supports our conjecture since the corresponding initial state factorizes. We claim that the competition between initial entanglement and dynamically created correlations is at the origin of the transition detected for the concurrence and the trace distance. The weaker the dynamics can entangle system and environment, the less the trace distance will be able to witness system-environment correlations as these correlations are either only slightly created in time or even completely destroyed. 

Although this explanation can also be applied to the swapped states $|\Psi^{\lambda}_s(0)\rangle$, it seems to fail for the states in the $\sigma_x$-eigenbasis \eqref{eq:initstatesigmax} as the concurrence always decreases from $[0,\pi]$. However, for $b_{1,2}=1/\sqrt{2}$, the total initial state can be rewritten as
\begin{equation}\label{eq:sigmax}
 |\Psi^{\lambda}_x(0)\rangle=\tfrac{1}{2}\Bigl\{|e\rangle\otimes(|0\rangle+|\Omega_\lambda\rangle)+|g\rangle\otimes(|0\rangle-|\Omega_\lambda\rangle)\Bigr\}\,,
\end{equation}
from which one directly observes that the states has vanishing coherences for any value of $\lambda$ as $|0\rangle+|\Omega_\lambda\rangle$ and $|0\rangle-|\Omega_\lambda\rangle$ are orthogonal. Clearly, for two-level states with fixed populations, states with zero coherences are most entangled. The states \eqref{eq:sigmax} thus initially saturate the upper bound for entanglement for the fixed populations determined by the value of $\lambda$. As the dynamics is a pure dephasing process with respect to the $\sigma_z$-basis, it introduces coherences and, therefore, reduces the amount of entanglement. The concurrence can thus only decrease.

\section{A spin star model}\label{sec:spinstar}
After having studied a two-level system coupled to a bosonic bath in terms of initial system-environment correlations, we now focus on a spin bath. We consider a system of $N+1$ spin-$\tfrac{1}{2}$ particles. One of the spins is located at the center and is coupled to the $N$ remaining spins, which are labeled by an index $k=1,...N$ with Pauli operators $\sigma^{(k)}$, via a Heisenberg $XY$ interaction \cite{hpbound,spinbath,bose}. This system is represented through the Hamiltonian
\begin{equation}
 H=A_0\sum_{k=1}^N(\sigma_+\sigma_-^{(k)}+\sigma_-\sigma_+^{(k)})\,,\label{eq:Hamiltonian Spin}
\end{equation}
 where $\sigma_\pm$ and $\sigma_\pm^{(k)}$ are the raising and lowering operators of the central spin and the $k$th bath spin, respectively. The Heisenberg $XY$ coupling has been found to yield an effective description for many physical systems such as quantum dots \cite{qdots}, and cavity QED \cite{qed} to name just a few.

 The dynamics of the central spin is exactly solvable so that we can study the detection of initial correlations with respect to the previously defined distance measures also for this model. For this purpose, we define an initially correlated total state similar to $|\Psi^{\lambda}_{U}(0)\rangle$ given in Eq.~\eqref{eq:initstate} and perform the same analysis as done for the single bosonic mode environment. The state reads
\begin{align}\label{eq:spininistate}
 |\xi^{\lambda}_{U}(0)\rangle=&b_1\bigl(u_{11} |e\rangle+u_{21}|g\rangle\bigr)\otimes |\chi_+\rangle \nonumber\\
&+b_2\bigl(u_{12} |e\rangle+u_{22}|g\rangle\bigr)\otimes |F_\lambda\rangle\,,
\end{align}
where
\begin{align}
 |F_\lambda\rangle=\tilde{C}_\lambda\bigl((1-\lambda)|\chi_+\rangle+\lambda|\chi_-\rangle\bigr)\,,
\end{align}
with $\tilde{C}_\lambda=1/\sqrt{\lambda^2+(1-\lambda)^2}$ ensuring the normalization of the state, and $|\chi_+\rangle=|\tfrac{N}{2},\tfrac{N}{2}\rangle$, $|\chi_-\rangle=|\tfrac{N}{2},\tfrac{N}{2}-1\rangle$ referring to two eigenstates of the total bath angular momentum $\vec{J}=\tfrac{1}{2}\sum_{k=1}^N \vec{\sigma}^{(k)}$ and its z-component $J_z$. As before, $|e\rangle$ and $|g\rangle$ denote the two levels of the open quantum system in its $\sigma_z$-basis. Finally, the constants $u_{ij}$ are the entries of the local unitary $U$ which solely acts on the open system Hilbert space.

\begin{figure}[!t]
   \centering
		\includegraphics[width=0.475\textwidth]{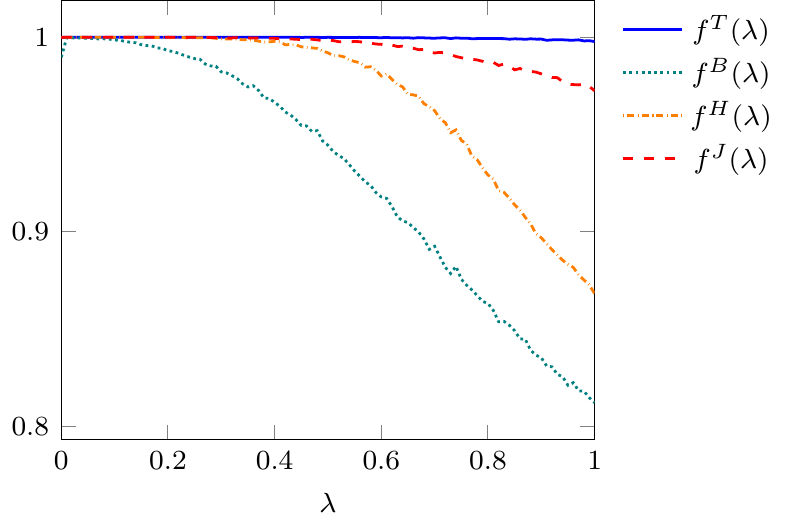}
    \caption{(Color online) Plot of the frequency of increase $f^k(\lambda)$ ($k=T,B,H,J$) of the considered distance measures for the marginals of the states $|\xi^{\lambda}_{U}(t)\rangle$ (cf. Eq.~\eqref{eq:spininistate}) with randomly sampled superposition amplitudes $b_{1,2}$ and local unitaries $U$ distributed with respect to the Haar measure. The sample size is $5\cdot10^4$.}\label{fig:spinstarunisampling}
 \end{figure}

The relative frequency of increase for the different measures is depicted in Fig.~\ref{fig:spinstarunisampling}\,. It shows that, also for the spin bath, the trace distance is more suited to detect the initial system-environment correlations present in the states $|\xi^{\lambda}_{U}(0)\rangle$ as it increases almost with certainty. However the difference between the considered distance measures is diminished. In comparison with the course of the frequency of increase for the Bures metric and the Hellinger distance for the first model, the evolution is reversed for the spin star model yielding a monotonically decreasing function of $\lambda$. That is, the stronger the correlations are, which are controlled by the parameter $\lambda$, the smaller is the capability of these two measures to increase above their respective initial values signifying the presence of initial correlations.

\section{Conclusions}\label{sec:conc}
In this work we studied the distance between quantum states in the presence of initial system-environment correlations with respect to four different distance measures: the trace distance, the Bures metric, the Hellinger distance and the Jensen-Shannon divergence. Correlations in quantum systems are an essential feature for many applications of quantum information processing schemes for which reason their detection is of particular interest. Our study performed on a two-level system which is coupled to a single mode environment, reveals fundamental differences pertaining to the qualitative behavior of the considered distance measures. Performing numerical studies we showed the distinguished rôle of the trace distance with respect to the capability of witnessing initial correlations, compared to the Bures metric and the Hellinger distance.

We illustrated  that the trace distance is far better suited to reliably detect initial correlations within the reduced dynamics for the considered model and, in particular, for the studied structure of the correlations than the three other measures. This is surprising as we could show that the Bures metric satisfies the same inequalities as the trace distance showing its potential capability of witnessing correlations in the same way $D_T$ does. This result is confirmed and strengthened by our findings for a spin star model on which we performed similar numerical studies. The outcome of our studies also suggests a certain relation between the trace distance and the Jensen-Shannon entropy, on the one hand, and the Bures metric and the Hellinger distance on the other hand.

In addition, we claim that there is a connection between the entangling capacity of the dynamics and the trace distance. When initial correlations forbid a further increase of entanglement quantified by the concurrence, the trace distance itself tends to decrease during the time evolution.  Hence, only for sufficiently weak correlations in the class of initial states \eqref{eq:initstatehan} the trace distance is capable to uncover the initial correlations. We remark that, since the total initial states are chosen to be pure, the correlations in these states are indeed fully quantum. The observed time evolution of the trace distance cannot be completely explained by the recently derived upper and lower bounds for the change of this distance measure \cite{laura1,laura2,andrea}. However, for sufficiently small correlations quantified by $\lambda$ these bounds provide some information on the dynamics of the trace distance.

The present work reveals a special status of the trace distance in comparison to the other measures. We conjecture that the relations found here are generic, showing that the trace distance is much more sensitive to initial system-environment correlations compared to the other measures. However, further studies on the features of these measures are required to give a complete and satisfying picture of their fundamental properties and differences.

\end{document}